\documentstyle[twoside,fleqn,espcrc2,epsf]{article}

\newcommand{\be}{\begin{equation}}
\newcommand{\ee}{\end{equation}}
\newcommand{\bdm}{\begin{displaymath}}
\newcommand{\edm}{\end{displaymath}}

\def\ham{{\bf \rm H}}

\def\bmat{{\bf \rm B}}
\def\cmat{{\bf \rm C}}

\def\spose#1{\hbox to 0pt{#1\hss}}
\def\ltapprox{\mathrel{\spose{\lower 3pt\hbox{$\mathchar"218$}}
 \raise 2.0pt\hbox{$\mathchar"13C$}}}
\def\gtapprox{\mathrel{\spose{\lower 3pt\hbox{$\mathchar"218$}}
 \raise 2.0pt\hbox{$\mathchar"13E$}}}
\def\inapprox{\mathrel{\spose{\lower 3pt\hbox{$\mathchar"218$}}
 \raise 2.0pt\hbox{$\mathchar"232$}}}
\title{ \hfill  FSU-SCRI-98C-88 \\
Spectrum of the Hermitian Wilson Dirac operator
}

\author{
Rajamani Narayanan
\address{
SCRI, The Florida State University, 
Tallahassee, FL 32306-4130, USA}
}

\begin{document}

\begin{abstract}

Recent results on the spectral properties of the Hermitian Wilson-Dirac
operator are presented.

\end{abstract}

\maketitle

% and here comes the text ...

\section{Introduction}

Continuum gauge field theory works under the assumption that all
fields are smooth functions of space time. This assumption is
certainly a valid one for quantum gauge field theories that respect
gauge invariance: One should always be able to fix a gauge so that the
gauge fields are smooth functions of space time since the action that
contains derivatives in gauge fields will not allow it otherwise. The
space of smooth gauge fields typically has an infinite number of
disconnected pieces with the number of pieces being in one to one
correspondence with the set of integers~\cite{topology}.
Every gauge field in each
piece can be smoothly interpolated to another gauge field in the same
piece but there is no smooth interpolation between gauge fields in
different pieces.  This is the case for U(1) gauge fields in two
dimensions and SU(N) gauge fields in four dimensions.

In lattice gauge theory, gauge fields are represented by link variables
$U_\mu(x)$ 
that are elements of the gauge group. Continuum derivatives are 
replaced by finite differences and the concept of smoothness of
gauge fields does not apply.
Any lattice gauge field configuration, $U_\mu(x)=e^{iA_\mu(x)}$
can be deformed to the trivial gauge field configuration by the
interpolation $U_\mu(x;\tau)=e^{i\tau A_\mu(x)}$ with 
$U_\mu(x;1)=U_\mu(x)$ and $U_\mu(x;0)=1$. Since smoothness does not
hold on the lattice away from the continuum limit, the space of gauge
fields on the lattice forms a simply connected space. Separation of the
gauge field space into an infinite number of disconnected pieces can only
be realized in the continuum limit. 

In this talk, I will address the following basic question: Do we see
a separation of lattice gauge fields configurations into topological
classes as we approach the continuum limit? To answer this question,
I will assume that an ensemble of lattice gauge field configurations
are given to me. This could be a pure gauge field ensemble or an
ensemble that includes fermion dynamics using either staggered fermions
or Wilson fermions. My aim is to answer this question without altering
the lattice gauge field configuration. I will use a Wilson-Dirac fermion
to {\sl probe} the lattice gauge field configuration. My motivation is
the overlap formalism~\cite{over} for chiral gauge theories. Topological aspects
of the background gauge field are properly realized by the chiral fermions
in this formalism and therefore it provides a good framework to answer
the above question. 
The hermitian Wilson Dirac operator enters 
the construction of lattice chiral fermions in the overlap formalism
and topological properties of the gauge fields are studied by looking
at the spectral flow of the hermitian Wilson Dirac operator as a 
function of the fermion mass. 

The talk is organized as follows. I begin by explaining the connection
between the spectral flow of the hermitian Wilson Dirac operator and
the topological content of the background gauge field. I then present
possible scenarios for the qualitative nature of the spectrum on the
lattice. I will then present some properties of the spectral flow,
and I will illustrate these properties by studying the spectral flow
in a single instanton background. I will then focus on the spectral
properties on lattice gauge field ensembles and their behavior as
the continuum limit is approached. I will only present an overview
of the results. Detailed results can be found in the talk presented
by Edwards~\cite{Robert}.

\section{Spectral flow, topology and condensates}

The massless Dirac operator in the continuum anticommutes with
$\gamma_5$. Therefore, the non-zero imaginary eigenvalues of
the massless Dirac operator come in pairs, $\pm i\lambda$ with
$\psi$ and $\gamma_5\psi$ being the two eigenvectors.
The zero eigenvalues of the massless Dirac operator are also
eigenvalues of $\gamma_5$. These chiral zero modes are a 
consequence of the topology of the background gauge field.
It is useful to consider the spectral flow of the Hermitian
Dirac operator:
\begin{equation}
\ham(m) = \gamma_5 (\gamma_\mu D_\mu+m )
\end{equation}
The non-zero eigenvalues of the massless Dirac operator
combine in pairs to give the following eigenvalue equation:
\begin{equation}
\ham(m) \chi_\pm = \pm \sqrt{\lambda^2+m^2} \chi_\pm
\end{equation}
%\begin{equation}
%\chi_+ = {1\over \sqrt{2}} \Bigl [ \psi + {m-i\lambda\over 
%\sqrt{\lambda^2 + m^2} }\gamma_5\psi \Bigr ]
%\end{equation}
%\begin{equation}
%\chi_- = {1\over \sqrt{2}} \Bigl [ \gamma_5\psi - {m+i\lambda\over 
%\sqrt{\lambda^2 + m^2} }\psi \Bigr ]
%\end{equation}
$\chi_\pm$ are linear combinations of $\psi$ and $\gamma_5\psi$ and
these modes never cross the x-axis in the spectral flow
of $\ham(m)$ as a function of $m$. 
The zero eigenvalues, $\gamma_\mu D_\mu\phi_\pm =0$ with
$ \gamma_5 \phi_\pm = \pm \phi_\pm$ result in
\begin{equation}
\ham(m) \phi_\pm = \pm m \phi_\pm
\end{equation}
These modes associated with topology result in flow lines
that cross the x-axis. A positive slope corresponds to positive
chirality and vice-versa. The net number of lines crossing zero
(the difference of positive and negative crossings) is the
topology of the background gauge field. 
Global topology of gauge fields cause exact zero eigenvalues at
$m=0$. In addition, one can have a non-zero spectral
density at zero. 
At infinite volume in the continuum, the spectrum is continuous and
$\rho(\lambda;m)d\lambda$ is the number of eigenvalues in the
infinitesimal region $d\lambda$. The spectral gap $\lambda_g(m)$
defined as the lowest eigenvalue at $m$
is equal to $|m|$. The spectral density at zero, $\rho(0;m)$ can be
non-zero only at $m=0$ indicating chiral symmetry breaking
in a theory like QCD. 
\begin{figure}
\epsfxsize=2.5in
%\centerline{\epsffile{spec_cont.ps}}
\centerline{\epsfbox[100 225 500 500]{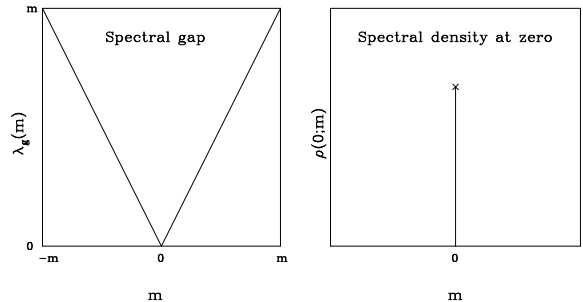}}
\caption{Continuum picture of the spectral gap and the spectral
density at zero.}
\label{fig:cont}
\end{figure}

To study the possible emergence of the above continuum picture 
as a continuum limit of a lattice gauge theory picture, we need
to have a lattice realization of $\ham(m)$. It is important to
note that we are interested in the spectral flow of a single
Dirac fermion. With this in mind, we choose the hermitian
Wilson-Dirac operator obtained by multiplying the standard Wilson-Dirac
operator by $\gamma_5$:
\begin{equation}
\ham_L(m) = 
\pmatrix {\bmat(U) +m & \cmat(U) \cr \cmat^\dagger(U) &  -\bmat(U) -m \cr}.
\end{equation}
$\cmat$ is the naive lattice first derivative term
and $\bmat$ is the Wilson term.
%\begin{eqnarray}
%\cmat_{i\alpha,j\beta}(k,k^\prime) \!\!\!\!\! &=& \!\!\!\!\! {1\over 2}
%\sum_\mu \sigma_\mu^{\alpha\beta}  \bigl[
%U^{ij}_\mu(k)\delta_{k^\prime,k+\hat\mu} \nonumber \\
%&-& (U^\dagger_\mu)^{ij}(k^\prime)
%\delta_{k,k^\prime+\hat\mu} 
%\bigr] 
%\label{eq:cmat}
%\end{eqnarray}
%is the naive discretization of the continuum Dirac operator and
%\begin{eqnarray}
%\bmat_{i\alpha,j\beta}(k,k^\prime) \!\!\!\!\! &=& \!\!\!\!\! 
%{1\over 2}\delta_{\alpha,\beta}
%\sum_\mu  \bigl[ 2\delta_{ij}\delta_{kk^\prime}-
%U^{ij}_\mu(k)\delta_{k^\prime,k+\hat\mu} \nonumber
%\\
%&-& (U^\dagger_\mu)^{ij}(k^\prime)
%\delta_{k,k^\prime+\hat\mu}\bigr]
%\quad
%\label{eq:bmat}
%\end{eqnarray}
%is the Wilson term that makes $\ham_L(m)$ describe a single massive 
%lattice fermion.
We are interested in the spectral flow of $\ham_L(m)$ as a function
of $m$. We note that 
$m=0,-2,-4,-6,-8$ are the points where the free fermions become massless
with degeneracies $1,4,6,4,1$ respectively. 
Next we observe that 
$\ham_L(m)$ can have a zero eigenvalue only if $m < 0$~\cite{over1}.
Let $\pmatrix{u\cr v\cr}$ be a normalized eigenvector of $\ham_L(m)$ with
zero eigenvalue. Then
$
mu+\bmat u + \cmat v =0$ and
$\cmat^\dagger u -\bmat v - mv =0$.
This implies that
$u^\dagger \bmat u + v^\dagger \bmat v = -m$.
Since $\bmat$ is a positive definite operator, the above equation can
have a solution only if $m < 0$.

\begin{figure}
\epsfxsize=2.5in
\centerline{\epsfbox[150 100 500 550]{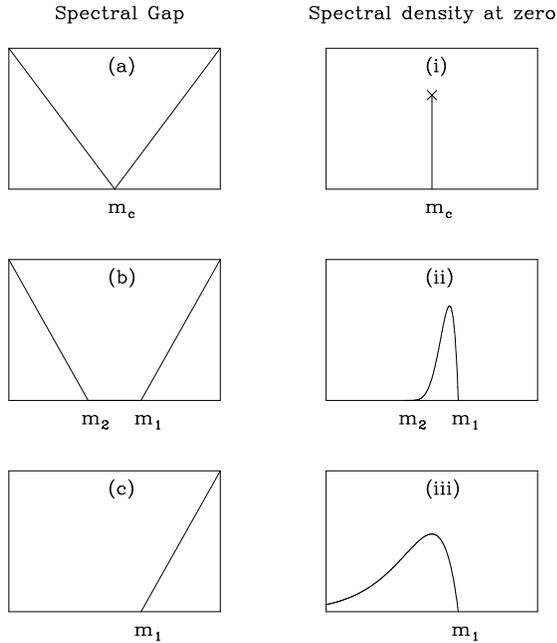}}
\caption{Possible scenarios of the spectral gap and the spectral
density at zero on the lattice.}
\label{fig:latt}
\end{figure}
We focus on the range $-2\le m \le 0$ and propose the following
scenarios for the spectral gap and the spectral density at zero
on the lattice and their approach to the continuum limit.
Six different scenarios are possible as shown in Fig.~\ref{fig:latt}.
\begin{itemize}
\item On the lattice we have (a) and (i) with $m_c\rightarrow 0$
in the continuum limit. $\rho(0;0)$ approaches the continuum
limit with proper scaling taken into account.
\item On the lattice we have (b) and (ii) with (a) and (i) being the
continuum limit. In this case,  $m_1\rightarrow 0$ and
$m_2\rightarrow 0$. In the limit we also get $\rho(0;0)$.
\item On the lattice we have (c) and (ii) with (a) and (i) being the
continuum limit. In this case,  $m_1\rightarrow 0$. The gap opens up
for $m < 0$ and again we get $\rho(0;0)$.
\item On the lattice we have (c) and (ii) with (c) and (i) being the
continuum limit. In this case,  $m_1\rightarrow 0$. But the gap 
does not open up
for $m < 0$.
\item On the lattice we have (c) and (iii) with (a) and (i) being the
continuum limit. In this case,  $m_1\rightarrow 0$. The gap opens up
for $m < 0$ and again we get $\rho(0;0)$.
\item {\it On the lattice we have (c) and (iii) with (c) and (i) being the
continuum limit. Here also  $m_1\rightarrow 0$
and $\rho(0,m)=0$ if $m< 0$. But the gap does not open up
for $m < 0$.}
\end{itemize}
I will show that numerical studies of the spectral flow on various
ensembles favor the last scenario. This seems to be the
case for any lattice
gauge ensemble whether the ensemble was generated with fermion
dynamics or not. One should keep in mind that the Wilson fermion
used to the study the spectral flow is only a probe and does not
enter the dynamics that generated the gauge field ensemble.

\section {Properties of the lattice spectral flow}

I now present a topological argument which will show that
zero eigenvalues of $\ham_L(m)$ can occur anywhere in
the region $-8 < m < 0$~\cite{smooth}. 
The spectrum of $\ham_L(m)$ and $-\ham_L(-8-m)$ are identical for
an arbitrary gauge field background. 
Since zero eigenvalues can occur only for $m<0$ in
$\ham_L(m)$, it follows that zero eigenvalues can occur only in
the region $-8 < m < 0$. It also follows that
every level crossing zero from above in the spectral flow of
$\ham_L(m)$ should be accompanied by a level crossing zero from below.
In a smooth instanton background a level crossing zero from above
at $m_1$ is accompanied by another level crossing zero from below
at $m_2 < m_1$. The second crossing is due to
one of the four doubler modes.
Both $m_1$ and $m_2$ will be functions of the size
of the instanton $\rho$ in lattice units. For $\rho >> a$, 
$m_1 \approx 0$ and $m_2 \approx -2$. As $\rho$ decreases,
$m_1$ moves farther away from zero and $m_2$ moves away from
$-2$ and closer to $m_1$. This motion as a function of $\rho$
is smooth and for some value of $\rho$, $m_1=m_2$.
Spectral flow changes smoothly as the configuration is
slowly changed. As we move in configuration space the topological
charge of a configuration changes. Tracing the spectral flow as
a function of configurations shows that
zero eigenvalues of $\ham_L(m)$ can occur anywhere in
the region $-8 < m < 0$.
\begin{figure}
\epsfxsize=2.in
\centerline{\epsfbox[150 100 500 550]{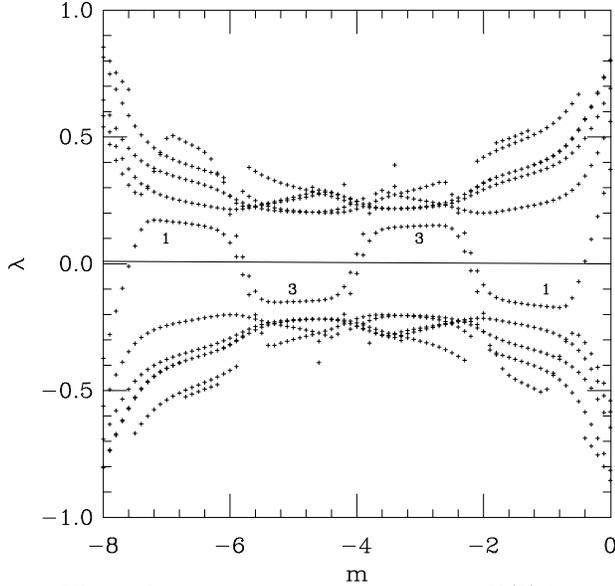}}
\caption{Spectral flow of a smooth SU(2) instanton with $\rho=2.0$ on
an $8^4$ lattice}
\label{fig:flow}
\end{figure}
\begin{figure}
\epsfxsize=2.in
\centerline{\epsfbox[150 100 500 550]{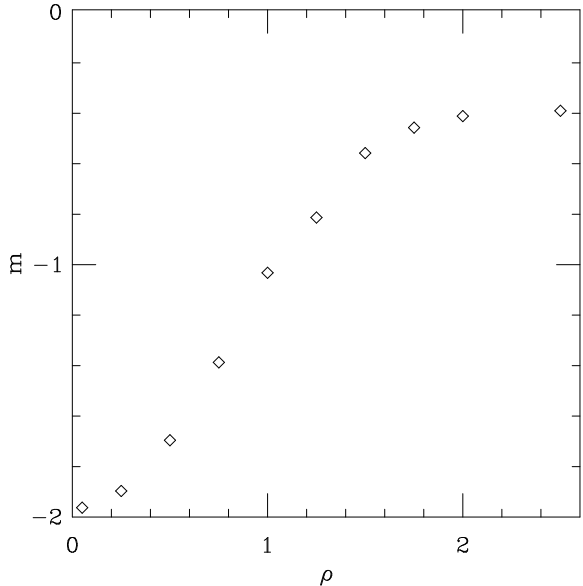}}
\caption{$m_1$ versus $\rho$ for a smooth instanton}
\label{fig:rho_vs_m}
\end{figure}

Fig.~\ref{fig:flow} illustrates the various properties of the spectral
flow of the Hermitian Wilson-Dirac operator. The background gauge field
is a smooth instanton with $\rho=2.0$ on a $8^4$ lattice. The first
crossing on the right of the figure from above to below is the one
associated with the particle feeling the global topology of the gauge field.
Then there are four levels crossing zero out of which three 
are degenerate.The degeneracy is a consequence of the smoothness of
the lattice instanton. The four crossings are attributed to the four
doublers feeling the global topology. This is followed by six levels
crossing zero, four levels crossing zero and finally one level crossing
zero. Clearly the spectrum of $\ham_L(m)$ and $-\ham_L(8-m)$ are the same.
There is no net level crossing in the range
$-8\le m \le 0$. This is  because the chiralities of the
sixteen particles are paired. Fig.~\ref{fig:rho_vs_m} shows the location
of the first level crossing zero as one varies the size of the instanton
in lattice units. As the size of the instanton is decreased the crossing
point moves farther away from zero. $\rho < 1$ instantons are placed
on the lattice by centering the instanton on a lattice site.
This figure shows that one can move the first crossing all the way up to
$m=-2$. 

\section{Spectral density at zero}

\begin{figure}
\epsfxsize=2.in
\centerline{\epsfbox[150 100 500 550]{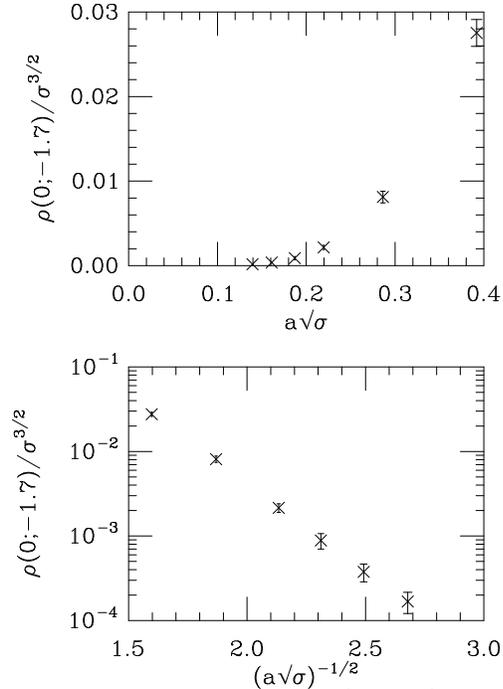}}
\caption{The approach of $\rho(0;-1.7)$ to the continuum limit}
\label{fig:rho0_scale_talk}
\end{figure}

In the previous section, I argued that $\ham_L(m)$ can have zero
crossings anywhere in the region $-2\le m \le 0$. Therefore the spectral
gap is zero in this region on the lattice. This has direct implications
for how the spectral density at zero behaves on the lattice.
A careful study of the spectral density at zero has been performed
on a variety of SU(3) pure gauge ensembles~\cite{su3_top}. 
This is done by computing
the low lying eigenvalues of $\ham_L(m)$ using the Ritz functional~\cite{Ritz}.
The low lying eigenvalues over the whole ensemble is then used to
obtain the integral of the spectral density function, namely
$\int_0^\lambda \rho(\lambda^\prime;m)d\lambda^\prime$. This
is then fitted to a polynomial and $\rho(0;m)$ is obtained as the 
coefficient of the linear term in the fit. 
\begin{figure}
\epsfxsize=2.in
\centerline{\epsfbox[100 100 500 420]{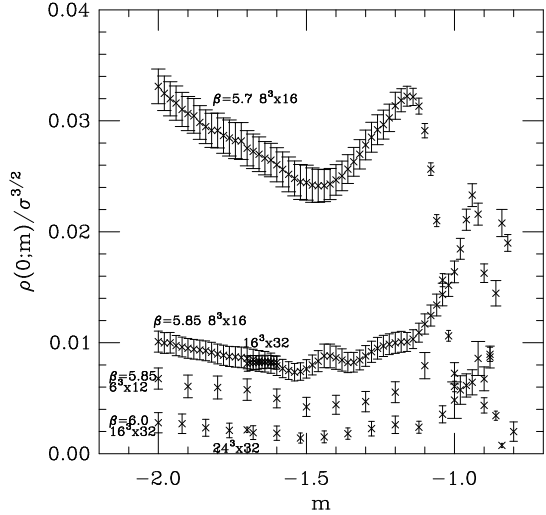}}
\caption{$\rho(0;m)$ as a function of $m$ for various SU(3) pure
gauge ensembles.}
\label{fig:rho0_talk}
\end{figure}
All ensembles show a peak in $\rho(0;m)$ at some value of $m$. There
is a sharp rise to the peak from the right and a gradual fall from 
the peak on the left. There is a gradual rise again to a second peak
at the location of the first set of doublers. 
The peak itself gets sharper as one goes toward the continuum limit.
It also moves to the right. $\rho(0;m)$ is non-zero for $-2 \le m \le m_1$
in the infinite volume limit at any finite value of the lattice gauge
coupling. $m_1$ goes to zero as the lattice coupling approaches the
continuum. $\rho(0;m)$ approaches the infinite lattice volume limit from
below as expected. We are fairly confident that we have
the infinite volume limit estimate for $\rho(0;m)$ 
at all the lattice spacings plotted in 
Fig.~\ref{fig:rho0_talk}. 
In Fig.~\ref{fig:rho0_scale_talk}, I focus on the behavior of
$\rho(0;m)$ at a fixed $m$ as one approaches the continuum limit.
The first figure indicates that $\rho(0;m)$ goes to zero
exponentially in the inverse
lattice spacing. This is given some credence by plotting the
same figure in a logarithmic scale. The function seems to favor
a fit of the form $be^{-c/\sqrt{a}}$. The power of $a$ in the exponent
is a consequence of an emperical fit but the data presents substantial
evidence for the following: $\rho(0;m)$ is non-zero for all finite lattice
spacings. The approach to zero at zero lattice spacing is faster than
any power of the lattice spacing. This shows that the last
scenario presented in the previous section is favored by the numerical
results on the lattice.

\section{Discussion}

A probe of lattice gauge fields using Wilson fermions has revealed that
fields are not continuum like on the lattice at gauge couplings
that are typically considered to be weak. If they were continuum like,
we should have seen evidence that $\rho(0,m)$ is non-zero
at a single value of $m$ or a region in $m$
that is of the order of the lattice spacing. Further we should have seen
a symmetry in the spectrum at values of $m$ on either side of the point
(or region) where $\rho(0,m)$ is non-zero. Instead, we found
that $\rho(0,m)$ is non-zero in a region $-2 \le m \le m_1$. In the
continuum limit, there is evidence that $m_1$ goes to zero and that
$\rho(0,m)$ goes to zero away from $m=0$. But the spectral distribution
does not show evidence for a symmetry as $m\rightarrow -m$. 

In addition to studying $\rho(0,m)$, one should also look at the
density of levels crossing zero in an infinitesimal range $dm$ centered
at $m$. In the continuum we expect level crossings only at $m=0$ but
we find a finite density of level crossings zero where ever $\rho(0;m)$
is non-zero~\cite{Robert}. 
This implies that the topological
charge of a lattice gauge field configuration defined as the net
level crossings in $\ham_L(m)$ in the range $[m,0]$ will depend on
$m$. The topology of a single lattice gauge field
configuration is not interesting in a field theoretic sense. One has
to obtain an ensemble average of the topological susceptibility and
study its dependence on $m$. This has been studied on a variety of
ensembles~\cite{Robert,su3_top}
 and the results show that the topological susceptibility is
essentially independent of $m$ in the region to the left of the peak
in $\rho(0;m)$. This indicates that the levels that cross zero to
the left of the peak are not physically relevant consistent with the
result that $\rho(0;m)$ goes to zero in the continuum limit.
The numbers for the topological susceptibility obtained in this manner
are consistent with the results obtained by field theoretic methods~\cite{topol}.
The modes that cross zero to the left of the peak in $\rho(0;m)$ correspond
to small modes~\cite{su3_top,Robert}
and this is consistent with Fig.~\ref{fig:rho_vs_m}.

A remark on the approach of $\rho(0,m)$ to the thermodynamic limit at
a fixed lattice gauge coupling is in order. For definiteness, let us
consider a pure gauge ensemble. If the extent of the lattice is
smaller than the extent needed for a finite temperature phase
transition then $\rho(0,m)$ will be essentially zero. One will see a
sharp increase in $\rho(0,m)$ at lattice volumes close to the phase
transition.  A table of $\beta_c(N_\tau)$ can be found
in~\cite{Fingberg} for SU(3) and SU(2).  For example, one needs to
have a lattice volume bigger than $8^4$ if one is using an SU(3) gauge
coupling of $\beta=6.0$ and one needs to work with a lattice larger
than $14^4$ at $\beta=6.38$.  Only then will one see a value of
$\rho(0,m)$ close to the thermodynamic limit.

Studies of the spectrum of the Wilson-Dirac operator is not in
any sense a new field of research. Smit and Vink~\cite{Vink}
studied the complex spectrum of the Wilson-Dirac operator to
understand topology. These initial studies led to the understanding
of dislocations in lattice field configurations that affect the
field theoretic determination of topological charge.
Spectral flows of the hermitian Wilson Dirac operator were
also studied to obtain some understanding of the role of topology
in the mesonic spectrum in quenched QCD~\cite{Yoshie}.
Distribution of real eigenvalues of the massless Wilson-Dirac operator
(identical to the zero eigenvalues of the Hermitian Wilson-Dirac
operator) has received a lot of attention recently 
motivated by the need to understand
the inability in extracting pion masses close to the chiral limit in
quenched QCD~\cite{recent}. 

\section*{Acknowledgments}

I would like to thank my collaborators, 
Robert Edwards and Urs Heller. All work I have done on
the spectrum of the Hermitian Wilson-Dirac operator
has been in collaboration with them. 
I would also like to thank Pavlos Vranas and Herbert Neuberger for
useful discussions.
This research was supported by DOE contracts 
DE-FG05-85ER250000 and DE-FG05-96ER40979.

% the bibliography comes at the end

\end{document}